\documentclass [hidelinks,12pt,openbib]{article}

\usepackage{cancel}
\usepackage{xcolor}
\usepackage[hidelinks]{hyperref}
\usepackage[normalem]{ulem}
 
\usepackage{mathtools}
 \usepackage{tikz}
 \usepackage{cancel}
\usepackage{fixmath}
\usepackage{dsfont}
\usepackage{caption}
\usepackage{pslatex}
\usepackage{bm}
\usepackage{graphicx}
\usepackage{amsmath,amssymb,amsfonts,amsthm}
\usepackage{xcolor}
\usepackage{mathtools}
\usepackage{dsfont}
\usepackage{cite}
\usepackage{ulem}
\usepackage{fancybox,fancyvrb}
\usepackage{framed} 

\usepackage{MnSymbol,wasysym}
\usepackage[export]{adjustbox}
\newtheorem{theorem}{Theorem}

\newtheorem{lemma}{Lemma}
\usepackage{enumitem}
\usepackage{color}
\usepackage{fancyhdr}
\usepackage{esint}
\usepackage{mathtools}
\usepackage[margin=1in]{geometry}
\usepackage{mdframed}

\def\XXint#1#2#3{{\setbox0=\hbox{$#1{#2#3}{\int}$}
     \vcenter{\hbox{$#2#3$}}\kern-.5\wd0}}
 \def\begfrwhite{\begin{mdframed}[
	backgroundcolor=white!15!white,
	linecolor=black,
	linewidth=0.5pt,
	align=center,
	userdefinedwidth=4.99in]}

\def\E1{\text{E}_1}

\def\L1loc{{L^1_{\rm loc}}}

\def\ig{\includegraphics}

\def\bc{\begin{center}}
\def\ec{\end{center}}

\def\begfr{\begin{mdframed}[
	backgroundcolor=gray!15!white,
	linecolor=blue,
	linewidth=1pt,
	align=center,
	userdefinedwidth=4.99in]}
\def\begfrwhite{\begin{mdframed}[
	backgroundcolor=white!15!white,
	linecolor=black,
	linewidth=0.5pt,
	align=center,
	userdefinedwidth=4.99in]}

\def\begfrblue{\begin{mdframed}[
	backgroundcolor=blue!15!white,
	linecolor=blue,
	linewidth=1pt,
	align=center,
	userdefinedwidth=4.99in]}

\def\endfr{\end{mdframed}}
\def\endfrred{\end{mdframed}}
\def\begfrred{\begin{mdframed}[
	backgroundcolor=red!15!white,
	linecolor=blue,
	linewidth=1pt,
	align=center,
	userdefinedwidth=4.99in]}
\def\endfrblue{\end{mdframed}}

\def\begeq{\begin{equation}}
\def\endeq{\end{equation}}

\def\bbr{\begin{bmatrix*}[r]}
\def\ebr{\end{bmatrix*}}
\def\bb{\begin{bmatrix*}[r]}
\def\eb{\end{bmatrix*}}
\def\bbc{\begin{bmatrix}}

\def\ebc{\end{bmatrix}}

\definecolor{darkmagenta}{rgb}{0.55, 0.0, 0.55}

\definecolor{darkgreen}{rgb}{0.0,0.6,0.0}

\def\R{\mathbb{R}}

\def\d{\displaystyle}

\def\0u{\underline{0}}

\def\begeq{\begin{equation}} 
\def\endeq{\end{equation}}

\def\bcenter{\begin{center}}
\def\ecenter{\end{center}}
\def\beq{\begin{equation}}
\def\eeq{\end{equation}}
\def\bmatr{\begin{bmatrix*}[r]}
\def\ematr{\end{bmatrix*}}
\def\bmatc{\begin{bmatrix}}
\def\ematc{\end{bmatrix}}

\usepackage{amsmath}               
  {
      \theoremstyle{plain}
      
  }
  {
      \theoremstyle{plain}
      
  }
   {
      \theoremstyle{plain}
      
  }
   {
      \theoremstyle{plain}
      
  }
  {
      \theoremstyle{plain}
      
  }
    {
      \theoremstyle{plain}
      
  }
\newcommand{\vertiii}[1]{{\left\vert\kern-0.25ex\left\vert\kern-0.25ex\left\vert #1 
    \right\vert\kern-0.25ex\right\vert\kern-0.25ex\right\vert}}
 \usepackage{amsthm}
\newtheorem*{proposition*}{Proposition}
\newtheorem*{corollary*}{Corollary}
\newtheorem*{definition*}{Definition}
\newtheorem*{example*}{Example}
\newtheorem*{lemma*}{Lemma}
\newtheorem*{theorem*}{Theorem}
\newtheorem*{observation*}{Observation}
\newtheorem*{remark*}{Remark}
\usepackage{sectsty}
\sectionfont{\fontsize{15}{10}\selectfont}

\newtheorem*{exmp*}{Example}
\usepackage{stmaryrd}
\usepackage{wrapfig}

\usepackage{xurl} 

\begin{document}

\begin{center}
{\bf \Large  Local wealth condensation for yard-sale models with \\ \vskip 5pt wealth-dependent biases} 
\vskip 15pt
Christoph B\"orgers$^1$ and Claude Greengard$^{2}$
\end{center}

\noindent
$^1$ Department of Mathematics, Tufts University, Medford, MA 

\noindent
$^2$ Two Sigma Investments, LP, New York, NY, and 

\noindent
\hskip 10pt Courant Institute of Mathematical Sciences, New York University, New York, NY

\vskip 20pt

\begin{quote}
{\small {\bf Abstract.} } In Chakraborti's {\em yard-sale model} of an economy \cite{original_yard_sale}, identical agents engage
 in pairwise trades, resulting in wealth exchanges
that conserve each agent's expected wealth.
Doob's martingale convergence theorem immediately implies almost sure {wealth condensation}, {\em i.e.}, convergence to a state in which a single agent owns the entire economy. If some pairs of agents are not allowed to trade with each other,
the martingale convergence theorem still implies {\em local} wealth condensation, {\em i.e.,} 
convergence to a state in which some agents are wealthy, while all their  trading partners are
impoverished. In this note, we  propose a new, more elementary proof of this result. Unlike the proof based on the martingale convergence theorem, our argument applies  to  models with a wealth-acquired advantage, 
and even to certain models with a  {\em poverty}-acquired advantage.
\end{quote}

\vskip 20pt
\section{Introduction} In Chakraborti's {\em yard-sale model}  \cite{original_yard_sale} of the economy, $N$ identical agents engage in pairwise trades.
After $\ell$ trades, the $i$-th agent owns the fraction $X_\ell^i  \in (0,1)$ of the total economy. We write
$$
X_\ell = \left[ X_\ell^i \right]_{1 \leq i \leq N}, ~~~ \ell=0,1,2,\ldots 
$$
and note that
$$
\sum_{i=1}^N X_\ell^i = 1 ~~~\mbox{for all $\ell$}.
$$

We think of $X_0$ as deterministic and given. The $X_\ell$ with $\ell \geq 1$ are random. In 
the original version of the model proposed by Chakraborti, they are defined inductively as follows.
Choose a number $b \in (0,1)$. Given $X_{\ell-1}$, choose a uniformly distributed random pair of integers
$$
 (i_\ell,j_\ell) \in \left\{(i,j) ~:~ 1 \leq i,j \leq N, ~ i \neq j \right\}, 
 $$ 
 and define $(\mu_\ell,\nu_\ell) = (i_\ell, j_\ell)$ if $X_{\ell-1}^{i_\ell} \leq X_{\ell-1}^{j_\ell}$, 
 and $(\mu_\ell,\nu_\ell) = (j_\ell, i_\ell)$ otherwise, so that in any case 
 $X_{\ell-1}^{\mu_\ell} \leq X_{\ell-1}^{\nu_\ell}$. Choose 
 a random $s_\ell \in \{-1,1\}$ with $P(s_\ell=1) = P(s_\ell=-1) = \frac{1}{2}$.   Assume that  $(i_\ell, j_\ell)$ and $s_\ell$ are independent of each other,
 and independent of the $X_k$, $(i_k,j_k)$, and $s_k$ with $k < \ell$.
Set 
 \begin{eqnarray}
 \label{eq:transfer} 
 &~& 
 X_\ell^{\mu_\ell} = X_{\ell-1}^{\mu_\ell} - s_\ell b X_{\ell-1}^{\mu_\ell}, ~~~
 X_\ell^{\nu_\ell} = X_{\ell-1}^{\nu_\ell} + s_\ell b  X_{\ell-1}^{\mu_\ell},
 \\
 \nonumber
&~& \mbox{and} ~
X_\ell^i = X_{\ell-1}^i ~\mbox{for $i \not \in \{\mu_\ell, \nu_\ell \}$.}
\end{eqnarray}
Thus a fraction $b$ of the poorer agent's wealth is transferred 
from one agent to the other, with the direction of transfer determined by 
a fair coin flip. 
This model is known to have the following striking property. 

\vskip 5pt

\begin{theorem} [Yard-Sale Convergence Theorem] 
\label{theorem:YST} 
The vectors $X_\ell$ converge to a canonical basis vector
of $\R^N$ almost surely.
\end{theorem} 

\vskip 5pt

In the limit, one agent comes to own everything. This sort of maximal inequality is called {\em wealth condensation}. 
In the yard-sale model, wealth condensation is the inescapable result of random, statistically unbiased interactions. Chakraborti first observed this  fact numerically \cite{original_yard_sale}.
For a version of the model in which there is a {continuum} of agents, rather than a finite number, an
analogous result was presented by Boghosian {\em et al} \cite{Boghosian_Johnson_Marcq}. 

Chorro 
 \cite{chorro} pointed out that Theorem \ref{theorem:YST}
  is an immediate consequence of Doob's martingale convergence theorem:
For a fixed $i$, the sequence $\{X_\ell^i \}_{\ell=0,1,2,\ldots}$ is a bounded martingale, and therefore must converge almost surely. It is
clear from the definition of the model that almost sure convergence of the
sequence  $\{X_\ell\}_{\ell=0,1,2,\ldots}$ implies almost sure convergence
to a canonical basis vector. Since $E(X_\ell^i) = E(X_0^i)$ for all $\ell$ and $i$, we also conclude
\begin{equation}
\label{eq:who_wins}
P \left( \lim_{\ell \rightarrow \infty} X_\ell^i = 1 \right) = X_0^i.
\end{equation}

Many variations can be analyzed similarly. For instance, the distribution 
of the pairs $(i_\ell, j_\ell)$  need not be uniform.  Different agents can be assumed
to have different degrees of
risk tolerance  \cite{cardoso_et_al_2023}, and their risk tolerance 
may even be history-dependent. Thus the number $b$ in \eqref{eq:transfer} might be replaced by random numbers
$B_\ell \in [\delta,1)$, where $\delta \in (0,1)$ is a fixed given number. The $B_\ell$ may depend on $X_0,\ldots,X_{\ell-1}$ and $(\mu_1,\nu_1), \ldots, (\mu_\ell, \nu_\ell)$. 
If $B_\ell$ is smaller, agent $\mu_\ell$ is more risk-averse during the $\ell$-th trade. 
We do need the assumption $B_\ell \geq \delta$, so 
that risk-taking never wholly disappears.
Obviously, but remarkably, eq.\ \eqref{eq:who_wins}  still holds for the modified model; risk aversion
does not affect an agent's likelihood of ending up owning the entire economy. 

\begin{figure}[h!] 
\bc
\ig[scale=0.9]{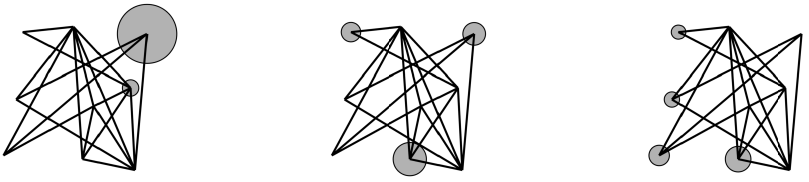}
\ec
\caption{Outcomes of three simulations of a yard-sale process on the same
 graph over many trades. The sizes of the
gray dots indicate the eventual wealth of the agents. There are two wealthy agents in the left panel, three in the middle panel, and four in the right panel.
The vertices representing wealthy agents are not directly connected by edges, so wealthy
agents never
trade with each other.} 
\label{fig:LOCALLY_RICH}
\end{figure}

An especially interesting variation is obtained when trades are allowed only among some, but 
not all, pairs of
agents \cite{Bustos_Guajardo_2012}. One can interpret this as an implementation of the model on an undirected
graph in which the vertices
are agents, and there is an edge between vertices $i$ and $j$ if and only if agents $i$ and $j$ are allowed 
to trade. In Chakrobarti's original model, the graph is {\em complete}, {\em i.e.}, any two vertices are
connected by an edge. When the graph is {\em incomplete}, 
there is {\em local} wealth condensation \cite{Bustos_Guajardo_2012,Lee_and_Lee_2023}; {\em i.e.}, if $a_{ij} \neq 0$, then 
$\min \{ X_\ell^i, X_\ell^j \} \rightarrow 0$ as $\ell \rightarrow \infty$. See Fig.\ \ref{fig:LOCALLY_RICH} for illustration.
This variation, too, can be analyzed easily using the martingale convergence theorem.

Cardoso {\em et al.}\ \cite{cardoso_et_al_2023} proposed an altogether different approach to proving
the yard-sale convergence theorem, based on the Gini index. Their analysis relies on what they call the  {\em fair rule hypothesis} \cite[Equation (8)]{cardoso_et_al_2023}. 
For wealth-conserving models, it is the martingale property.  
Most, though not all, examples in \cite{cardoso_et_al_2023} are wealth-conserving. The main result in \cite{cardoso_et_al_2023} is that the Gini index is monotonically
increasing, and stationary if and only if it assumes its maximal value.

Arguments based on the martingale property break down when the coin flips governing who benefits from a trade
are taken to be biased. 
Boghosian {\em et al.}\ \cite{Boghosian_and_gang} allowed
a {wealth-acquired advantage}: 
$s_\ell = 1$ with some  probability $p \in  \left[\frac{1}{2}, 1 \right)$. 
One should certainly expect wealth condensation for $p>\frac{1}{2}$, considering
that it occurs even for $p= \frac{1}{2}$. 
More surprisingly, Moukarzel {\em et al.} \cite{Moukarzel_et_al_2007} gave a heuristic
argument showing that even with a very slight {\em poverty}-acquired advantage, there can be wealth condensation. 
In these models, $\{X_\ell^i\}_{\ell \geq 0}$ is  no longer
a martingale, nor a sub- or super-martingale. 

In this paper, which generalizes our preprint \cite{Borgers_Greengard_2023}, we give a new probabilistic analysis of the results sketched
above, not relying on the martingale property. We
use  $\| X_\ell \|^2$ as a measure of concentration, and analyze its time evolution. 
For any non-zero vector $X \in \R^N$, the ratio $\| X \|_2/\| X \|_1$ is $\leq 1$, and can
be viewed as a measure of concentration. In fact, this ratio is equal to 1 if and only if $X$ is ``maximally concentrated," namely, a non-zero multiple of a 
canonical basis vector.
This observation has found
many uses, for instance  in 
quantum physics \cite{inverse_participation_ratio}, political science \cite{Laakso_Taagepera}, ecology \cite{simpson_diversity}, and   antitrust regulation \cite{justice_department}. It also underlies the idea
of  adding an $L^1$ penalty term in regression, the ``lasso" method \cite{lasso}, to obtain 
sparse (namely, concentrated) solutions. 

\vskip 10pt

\section{Mathematical statement of our model} We will state our precise
mathematical assumptions without repeating the
economic motivation already given in the introduction. 
We assume that we are given
\vskip 5pt
\begin{enumerate}
\item[-] an integer $N \geq 2$, 
\item[-] a vector $X_0 \in (0,1)^N$ with $\sum_{i=1}^N X_0^i = 1$, 
\item[-] an adjacency matrix $A = \left[ a_{ij} \right]_{1 \leq i, j \leq N}$ with $a_{ii}=0$ for all $i$, $a_{ij} = a_{ji} \in \{0,1\}$ for $i \neq j$, and $a_{ij} = a_{ji} = 1$ for least one pair $(i,j)$, 
\item[-] a probability distribution $\pi$ on the set
$$
{\cal E} = \left\{(i,j) ~:~ 1 \leq i,j \leq N, ~ a_{ij} = 1 \right\}
$$
assigning to each element of ${\cal E}$ a positive probability, 
\item[-] a sequence of pairs $(i_\ell, j_\ell)$, $\ell \geq 1$, independent of each other with probability
distribution $\pi$, 
\item[-] sequences of random numbers $\{ U_\ell \}_{\ell \geq 1}$, 
$\{ V_\ell \}_{\ell \geq 1}$, and $\{ W_\ell \}_{\ell \geq 1}$, uniformly distributed in $(0,1)$ and 
independent of each other and of the 
pairs $(i_\ell, j_\ell)$, 
\item[-] Borel measurable functions
$$
f_\ell: ~ \left( \R^N \right)^\ell  \times {\cal E}^\ell  \times (0,1)  \rightarrow [\delta, 1), ~~~~~\ell \geq 1, 
$$
where $\delta \in (0,1)$ is a fixed given number, independent of $\ell$, 
\item[-] Borel measurable functions 
$$
g_\ell: ~ \left( \R^N \right)^\ell  \times {\cal E}^\ell  \times (0,1)  \rightarrow (0, 1), ~~~~~\ell \geq 1. 
$$
\end{enumerate}

Given these data, we define random
vectors $X_\ell \in (0,1)^N$ with $\sum_{i=1}^N X_\ell^i=1$, $\ell=1,2,\ldots$, inductively as follows.
Given $X_{\ell-1}$, set $(\mu_\ell, \nu_\ell) = (i_\ell, j_\ell)$ if $X_{\ell-1}^{i_\ell} \leq X_{\ell-1}^{j_\ell}$, and 
$(\mu_\ell, \nu_\ell) = (j_\ell, i_\ell)$ otherwise.  In order to allow history dependence of $B_\ell$ and $p_\ell$, yet leave room for randomness, we define
\begin{equation}
\label{eq:f_ell}
B_\ell = f_\ell(X_0,X_1,\ldots,X_{\ell-1},i_1,j_1,i_2,j_2, \ldots, i_\ell, j_\ell, U_\ell)  \in [\delta, 1) 
\end{equation}
and 
\begin{equation}
\label{eq:g_ell}
p_\ell = g_\ell(X_0,X_1,\ldots,X_{\ell-1},i_1,j_1,i_2,j_2, \ldots, i_\ell, j_\ell, V_\ell)  \in (0,1). 
\end{equation}
The assumption that $U_\ell$ and $V_\ell$ in eqs.\ 
\eqref{eq:f_ell} and \eqref{eq:g_ell} are uniformly distributed is not restrictive, since
for any distribution $\Phi$ on $\R$, there exists a Borel-measurable function 
$\varphi$ so that $\varphi(Z)$ has 
distribution $\Phi$ if $Z \in (0,1)$ is uniformly
distributed \cite[Theorem 1.2.2]{Durrett_book}. 
Let further 
$$
s_\ell = \left\{ \begin{array}{rl} 1 & \mbox{if $W_\ell \leq p_\ell$}, \\
-1 & \mbox{otherwise}.
\end{array}
\right.
$$
Then define 
 \begin{eqnarray}
 \label{eq:trade} 
 &~& 
 X_\ell^{\mu_\ell} = X_{\ell-1}^{\mu_\ell} - s_\ell B_\ell X_{\ell-1}^{\mu_\ell}, ~~~
 X_\ell^{\nu_\ell} = X_{\ell-1}^{\nu_\ell} + s_\ell B_\ell  X_{\ell-1}^{\mu_\ell},
 \\
 \nonumber
&~& \mbox{and} ~
X_\ell^i = X_{\ell-1}^i ~\mbox{for $i \not \in \{\mu_\ell, \nu_\ell \}$.}
\end{eqnarray}

\vskip 10pt

\section{Conditional expected change in the Euclidean norm of the \\ \vskip -3pt 
\hskip 6pt wealth vector in a single trade}
The key step in our analysis is to calculate the conditional expected change in the Euclidean norm
of the vector $X_\ell$ in one trade: 
\begin{eqnarray}
\nonumber
&~& E \left( \| X_\ell \|^2 - \|X_{\ell-1}\|^2 ~|~ X_0,\ldots,X_{\ell-1}, \mu_\ell, \nu_\ell, B_\ell, p_\ell \right)  \\
\nonumber
&=& p_\ell   \left( \left( X_{\ell-1}^{\mu_\ell} - B_\ell X_{\ell-1}^{\mu_\ell} \right)^2 + \left( 
X_{\ell-1}^{\nu_\ell} + B_\ell  X_{\ell-1}^{\mu_\ell} \right)^2 \right)  +  \\
\nonumber
&~& (1-p_\ell)  \left( \left( X_{\ell-1}^{\mu_\ell} + B_\ell X_{\ell-1}^{\mu_\ell} \right)^2 + \left( 
X_{\ell-1}^{\nu_\ell} -  B_\ell  X_{\ell-1}^{\mu_\ell} \right)^2  \right) - \left(  \left( X_{\ell-1}^{\mu_\ell} \right)^2 + \left( X_{\ell-1}^{\nu_\ell} \right)^2  \right)  \\
\label{eq:condE} 
&=& (4 p_\ell-2)  B_\ell X_{\ell-1}^{\mu_\ell} \left( X_{\ell-1}^{\nu_\ell} - X_{\ell-1}^{\mu_\ell} \right) + 2\left( B_\ell X_{\ell-1}^{\mu_\ell} \right)^2 .
\end{eqnarray}
For fair coin tosses ($p_\ell=\frac{1}{2}$ for all $\ell$), the left summand in \eqref{eq:condE} is zero, 
and therefore \eqref{eq:condE} is positive. 
For coin tosses with a wealth-acquired advantage  ($p_\ell>\frac{1}{2}$), \eqref{eq:condE} is 
of course positive as well. In order to demonstrate condensation even 
in some cases with 
a poverty-acquired advantage ($p_\ell<\frac{1}{2}$), we assume
\begin{equation}
\label{eq:p_condition}
(4 p_\ell-2)   \left( X_{\ell-1}^{\nu_\ell} - X_{\ell-1}^{\mu_\ell} \right) \geq - \delta X_{\ell-1}^{\mu_\ell} ~~~\mbox{almost surely}. 
\end{equation}
Note that this is an assumption on the functions $g_\ell$. Since $-\delta X_{\ell-1}^{\mu_\ell} \geq - B_\ell X_{\ell-1}^{\mu_\ell}$, 
it then follows that \eqref{eq:condE} is $\geq \left( B_\ell X_{\ell-1}^{\mu_\ell} \right)^2$.
We summarize our result in the following lemma. 

\vskip 5pt
\begin{lemma} Inequality \eqref{eq:p_condition} implies 
\begin{equation}
\label{eq:cond_E_estimate}
E \left( \| X_\ell \|^2 - \|X_{\ell-1}\|^2 ~|~ X_0,\ldots,X_{\ell-1}, \mu_\ell, \nu_\ell, B_\ell, p_\ell \right) \geq 
\left( B_\ell X_{\ell-1}^{\mu_\ell} \right)^2.
\end{equation}
\end{lemma}

\vskip 20pt

\section{Convergence of the amount of wealth at stake in a transaction}

\begin{lemma} 
\label{prop:wealth_at_stake}
Inequality \eqref{eq:p_condition} implies 
 $\d{\lim_{\ell \rightarrow \infty} X_{\ell-1}^{\mu_\ell}= 0} $ almost surely.  
\end{lemma}

\vskip 5pt
\begin{proof} 
Let $\epsilon>0$. We will show that almost surely, $(B_\ell X_{\ell-1}^{\mu_\ell} )^2 \geq \epsilon$
for at most finitely many $n$. By the Borel-Cantelli lemma, it is sufficient to show that 
$$
\sum_{\ell=1}^\infty P \left( (B_\ell X_{\ell-1}^{\mu_\ell} )^2 \geq \epsilon \right) < \infty.
$$
Since 
$$
E \left( \left( B_\ell X_{\ell-1}^{\mu_\ell} \right)^2 \right) \geq \epsilon^2 P \left( (B_\ell X_{\ell-1}^{\mu_\ell} )^2 \geq \epsilon \right),
$$
it suffices to prove 
$$
\sum_{\ell=1}^\infty E \left( (B_\ell X_{\ell-1}^{\mu_\ell} )^2  \right) < \infty.
$$
Taking (unconditional) expectations in \eqref{eq:cond_E_estimate}, we find 
$$
E((B_\ell X_{\ell-1}^{\mu_\ell})^2) \leq E(\| X_\ell\|^2 - \| X_{\ell-1} \|^2).
$$
and therefore, for all $k$, 
$$
\sum_{\ell=1}^k E \left( (B_\ell X_{\ell-1}^{\mu_\ell} )^2  \right) \leq E(\| X_k \|^2) - E(\| X_0 \|^2) \leq 
E \left( \sum_{i=1}^N \left( X_k^i \right)^2 \right)   \leq E \left(  \sum_{i=1}^N X_k^i  \right) =1.
$$

We conclude that $(B_\ell X_{\ell-1}^{\mu_\ell})^2 \rightarrow 0$ almost surely, and since
$B_\ell \geq \delta>0$, this implies $X_{\ell-1}^{\mu_\ell} \rightarrow 0$ almost surely.
\end{proof} 

\vskip 10pt
\section{Local wealth condensation} We write 
$$
{\cal C} = \left\{ X \in [0,1]^N ~:~ \sum_{i=1}^N X^i = 1, ~~  X^i X^j a_{ij} = 0 ~\mbox{for all $(i,j)$ with 
$1 \leq i, j \leq N$} \right\}.
$$
A wealth distribution in ${\cal C}$ has the property that any agent with a positive 
amount of wealth is connected only to agents with zero wealth. Local wealth condensation
means convergence of the $X_\ell$ to 
${\cal C}$.

\vskip 5pt
\begin{theorem}[Generalized Yard-Sale Convergence Theorem]
Inequality \eqref{eq:p_condition} implies 
\begin{equation}
\label{eq:local}
\lim_{\ell \rightarrow \infty} {\rm dist}(X_\ell, {\cal C}) = 0 ~~~~\mbox{almost surely}. 
\end{equation}
Here {\rm dist} denotes the distance in any norm on $\R^N$.
\end{theorem}

\vskip 5pt
\begin{proof}
Equation \eqref{eq:local} means
$$
\lim_{\ell \rightarrow \infty} 
\max \left\{ \min \left\{ X_{\ell-1}^i , X_{\ell-1}^j \right\} ~:~ 1 \leq i, j \leq N, ~ a_{ij} = 1 \right\}=0 ~~~
\mbox{almost surely}. 
$$
If this were not the case, then with positive probability, there would be an $\epsilon>0$ so that 
$$
\max \left\{  \min \left\{ X_{\ell-1}^i , X_{\ell-1}^j \right\}~:~ 1 \leq i, j \leq N, ~ a_{ij} = 1 \right\} \geq \epsilon.
$$
infinitely often. This would imply that with positive probability, $X_{\ell-1}^{\mu_\ell} \geq \epsilon$ infinitely often, in contradiction with 
Lemma \ref{prop:wealth_at_stake}.
\end{proof}

\vskip 10pt

\section{Discussion} 
\label{sec:main}
The yard-sale convergence theorem shows that (extreme) inequality can be the result of pure
chance, and does not require agents of varying ability or industriousness. This is why it is 
interesting. 

In this paper, we have tried to strike a balance between clarity and generality.  
We could, for instance, have allowed the choice of $(i_\ell, j_\ell)$ to be history-dependent, as long as for
every $(i,j) \in {\cal E}$, the probability $P\left( \left( i_\ell, j_\ell \right) = \left( i, j \right) \right)$ is uniformly
bounded away from zero, so that no pair of agents drops out altogether in the limit as $\ell \rightarrow \infty$.
Further, instead 
of $B_\ell \geq \delta$, we could have assumed that $P \left( B_\ell \geq \delta \right)$ is uniformly bounded away from zero. 
These generalizations didn't seem to add insight justifying 
the expense of more opaque notation.

Numerical experiments and heuristic arguments \cite{Moukarzel_et_al_2007} suggest
that wealth condensation even occurs for a fixed $p_\ell=p$ very slightly smaller than $\frac{1}{2}$.
However,  \eqref{eq:p_condition} does not hold then, so our theorem does not extend to this
case. 

It is plausible and supported by numerical simulations that  eq.\ \eqref{eq:local} can be replaced by the slightly stronger statement
\begin{equation}
\label{eq:local_strong}
\exists X \in {\cal C} ~~~ \lim_{\ell \rightarrow \infty} X_\ell = X.
\end{equation}
For a  complete graph, ${\cal C}$ consists of isolated points only, namely,  the canonical basis vectors in $\R^N$. Conditions
\eqref{eq:local} and \eqref{eq:local_strong} are  equivalent in that case. 
For an incomplete graph, ${\cal C}$ always
contains non-isolated points, and the equivalence of \eqref{eq:local} and \eqref{eq:local_strong} is therefore no longer  obvious. However, for $p=1/2$, the argument based on the martingale convergence
theorem does prove \eqref{eq:local_strong} even for an incomplete graph.

These limitations notwithstanding, the argument given in this paper appears
to be both the simplest and the most general rigorous mathematical proof of the yard-sale convergence
theorem for a finite number of agents.


\begin{thebibliography}{10}

\bibitem{Boghosian_and_gang}
{\sc B.~Boghosian, A.~Devitt-Lee, M.~Johnson, J.~Marcq, and H.~Wang}, {\em
  {Oligarchy as a phase transition: The effect of wealth-attained advantage in
  a Fokker-Planck description of asset exchange}}, Physica A, 476 (2017),
  pp.~15--37.

\bibitem{Boghosian_Johnson_Marcq}
{\sc B.~Boghosian, M.~Johnson, and J.~Marcq}, {\em {An H theorem for
  Boltzmann's equation for the yard-sale model of asset exchange}}, J. Stat.
  Phys., 161 (2015), pp.~1339--1350.

\bibitem{Borgers_Greengard_2023}
{\sc C.~B{\"o}rgers and C.~Greengard}, {\em A new probabilistic analysis of the
  yard-sale model}, ArXiv e-print 2308.01485v1, 2023.

\bibitem{Bustos_Guajardo_2012}
{\sc R.~Bustos-Guajardo and C.~F. Moukarzel}, {\em Yard-sale exchange on
  networks: wealth sharing and wealth appropriation}, J. Stat. Mech.: Theory
  Exp., {\rm P12009},  (2012).

\bibitem{cardoso_et_al_2023}
{\sc B.-H.~F. Cardoso, S.~Gon\c{c}alves, and J.~R. Iglesias}, {\em {Why equal
  opportunities lead to maximum inequality? The wealth condensation paradox
  generally solved}}, Chaos Solit. Fractals, 168, 113181 (2023).

\bibitem{original_yard_sale}
{\sc A.~Chakraborti}, {\em {Distributions of money in model markets of
  economy}}, Int. J. Mod. Phs. C, 13 (2002), pp.~1315--1321.

\bibitem{chorro}
{\sc C.~Chorro}, {\em A simple probabilistic approach of the yard-sale model},
  Stat.\ Probab.\ Lett., 112 (2016), pp.~35--40.

\bibitem{Durrett_book}
{\sc R.~Durrett}, {\em Probability: Theory and Examples}, Cambridge University
  Press, 2019.

\bibitem{inverse_participation_ratio}
{\sc B.~Kramer and A.~MacKinnon}, {\em {Localization: theory and experiment}},
  Rep. Prog. Phys., 56 (1993), pp.~1469--1564.

\bibitem{Laakso_Taagepera}
{\sc M.~Laakso and R.~Taagepera}, {\em {``Effective" number of parties: a
  measure with applications to West Europe}}, Comp. Polit. Stud., 12 (1979),
  pp.~3--27.

\bibitem{Lee_and_Lee_2023}
{\sc H.~G. Lee and D.-S. Lee}, {\em Scaling in local to global condensation of
  wealth on sparse networks}, Phys. Rev. E, 108, doi:
  10.1103/PhysRevE.108.064303 (2023).

\bibitem{Moukarzel_et_al_2007}
{\sc C.~F. Moukarzel, S.~Con\c{c}alves, J.~R. Iglesias, M.~Rodr\'iguez-Achach,
  and R.~Huerta-Quintanilla}, {\em Wealth condensation in a multiplicative
  random asset exchange model}, Eur. Phys. J. Special Topics, 143 (2007),
  pp.~75--79.

\bibitem{simpson_diversity}
{\sc E.~H. Simpson}, {\em {Measurement of diversity}}, Nature, 163 (1949).

\bibitem{lasso}
{\sc R.\ Tibshirani}, {\em Regression shrinkage and selection via the lasso},
  J.\ Royal Stat.\ Soc.\ B, 58 (1996), pp.~267--288.

\bibitem{justice_department}
{\sc {U.S. Department of Justice}}, {\em {Herfindahl-Hirschmann
  Index,}\hskip 75pt}
\newblock https://www.justice.gov/atr/herfindahl-hirschman-index.

\end{thebibliography}

 \end{document}